\documentclass
[superscriptaddress,secnumarabic,amssymb,amsmath,nobibnotes,aps,prd,showkeys,showpacs,twocolumn,nofootinbib]{revtex4}%
\usepackage{graphicx}
\usepackage{xcolor}
\usepackage{epsf}
\usepackage{bm}
\usepackage{amsmath}
\usepackage{amsfonts}
\usepackage{amssymb}
\usepackage{epstopdf}%
\providecommand{\U}[1]{\protect\rule{.1in}{.1in}}
\newcommand{\be}{\begin{equation}}
\newcommand{\ee}{\end{equation}}

\newcommand{\mincir}{\raise
-3.truept\hbox{\rlap{\hbox{$\sim$}}\raise4.truept\hbox{$<$}\ }}
\newcommand{\magcir}{\raise
-3.truept\hbox{\rlap{\hbox{$\sim$}}\raise4.truept\hbox{$>$}\ }}

\begin{document}
\title{The Palatini star: exact solutions of the modified Lane-Emden equation}

\author{Artur Sergyeyev}
\email{artur.sergyeyev@math.slu.cz}
\affiliation{Mathematical Institute, Silesian University in Opava, Na Rybn\'\i{}\v{c}ku 1, 74601 Opava, Czech Republic}

\author{Aneta  Wojnar}
\email{aneta.wojnar@cosmo-ufes.org}
\affiliation{N{\'u}cleo Cosmo-ufes \& PPGCosmo, Universidade Federal do Esp{\'i}rito Santo,
29075-910, Vit{\'o}ria, ES, Brasil.}


\date{\today}

\begin{abstract}
Two exact solutions for $n=0$ and $n=1$ of the Palatini-modified Lane-Emden equation are found. We have employed these solutions to describe a Palatini-Newtonian neutron star and compared the result with the pure Newtonian counterpart. It turned out that for the negative parameter of the Starobinsky model the star is heavier and larger.
\end{abstract}

\maketitle

\section{Introduction}
Although there is no doubt of the beauty and the overall  correctness of the General Relativity \cite{ein1,ein2}, the theory suffers from a number of shortcomings \cite{Copeland:2006wr,Nojiri:2006ri,Capozziello:2007ec,Carroll:2004de,Sotiriou:2008ve} which could be addressed by extensions, provided by the Extended Theories of Gravity (ETGs), of the Einstein's proposal.
Apart from the cosmological arguments
on the need of looking for modified theories of gravity
\cite{Cap_Laur,Cap_Far,Starobinsky:1980te,Guth:1980zm,Huterer:1998qv,Sami},
there are also indications coming from astrophysics strengthening reasons for searching
for the alternatives. In the light of the very recent observation of the pulsar PSR J2215+5135 with the mass $M_1=2.27^{+0.17}_{-0.15}M_\odot$ for the 
neutron star \cite{lina} putting in trouble exotic forms of matter (hyperons or deconfined quarks) and the previous discoveries of the massive neutron stars as $2.01\pm0.04M_\odot$ (PSR J0348+0432) \cite{as}, and $1.97\pm0.04M_\odot$ 
(PSR J1614-2230) \cite{craw}, the ETGs \cite{Cap_beyond, cap_invar} allow to exceed the maximal mass of $2M_\odot$ demonstrated by GR without introducing exotic components. 

Roughly speaking, in the ETGs one changes the geometric part of the action which can be achieved by, for example, generalizing the Einstein-Hilbert action Lagrangian density to an arbitrary function  of the scalar curvature ($f(R)$ gravity \cite{Starobinsky:1980te, Sotiriou:2008ve,buchdahl,cap_jcap,cap_not,DeFelice:2010aj,Will:1993te,Palatini:1919di,Capozziello:2011et,Ferraris:1992dx}), by adding minimally or non-minimally coupled scalar field \cite{brans, Bergmann}, or by assuming that physical constants are not actually constants  \cite{Dabrowski:2012eb, Leszczynska:2014xba, Salzano:2016pny}.

In the present paper we are interested in the $f(\hat{R})$ Palatini gravity under the Ehlers-Pirani-Schild (EPS) interpretation, which in more detail was
discussed in \cite{eps, mauro, fatibene}. The Palatini formulation of $f(\hat R)$ gravity has  advantages with respect to the metric one since the field equation are second-order differential equations for the metric $g$ while in the metric formulation we deal with the fourth-order equations. Since we are interested in stars, it should be mentioned that astrophysical objects have been already widely discussed in the context of Palatini gravity: studies on black holes can be found in 
\cite{diego1,diego2,diego3,diego4,diego5,diego6,man}, on wormholes in \cite{worm, worm2, diego7,diego8}, and on neutron stars in 
\cite{kain,reij,pano,anab, anet,Barausse:2007ys, barau, barau2, pani, sham, olmo_ns, mana}.

Let us just briefly recall the Palatini formalism which assumes that the spacetime geometry is described by two independent structures: a class of Lorentzian metrics which are conformally related to each other, and a connection which later on, as a dynamical feature of the action, will turn out to be a Levi-Civita connection with respect to a metric conformally related to the metric $g$. Thus, we consider a standard action of the $f(\hat R)$ gravity of the form
\begin{equation}
S=S_{\text{g}}+S_{\text{m}}=\frac{1}{2\kappa}\int \sqrt{-g}f(\hat{R}) d^4 x+S_{\text{m}}(g_{\mu\nu},\psi_m),\label{action}
\end{equation}
where $\hat{R}=\hat{R}^{\mu\nu}g_{\mu\nu}$ is the Palatini-Ricci scalar constructed using two independent objects, the connection $\tilde\Gamma$ and the metric $g$. The variation with respect to the metric $g$ and the independent connection $\tilde\Gamma$ is given, respectively, by the formulas
\begin{align}
   f'(\hat{R})\hat{R}_{\mu\nu}-\frac{1}{2}f(\hat{R})g_{\mu\nu}=&\kappa T_{\mu\nu},\label{structural}\\
   \hat{\nabla}_\beta(\sqrt{-g}f'(\hat{R})g^{\mu\nu})=&0,\label{con}
\end{align}
where $T_{\mu\nu}$ is the energy momentum tensor and $f'(\hat{R})=df/d\hat{R}$. 

Equations (\ref{con}) imply that the connection $\hat{\nabla}_\beta$ is the Levi-Civita connection of the conformal metric $h$:
\begin{equation}
    h_{\mu\nu}=f'(\hat{R})g_{\mu\nu}.
\end{equation}
Let us observe that Palatini gravity is equivalent to GR when one considers the linear function $f$ of the form $f(\hat R)=\hat R-2\Lambda$, and the independent connection turns out to be Levi-Civita connection of the metric $g$.

We are going to use here the following interesting fact: the equations in question can be rewritten  \cite{DeFelice:2010aj} in the terms of the conformal metric $ h_{\mu\nu}$ \cite{BSS,SSB} and the scalar 
 field defined as $\Phi=f'(\hat{R})$:
 \begin{subequations}
	\begin{align}
	\label{EOM_P1}
	 \bar R_{\mu\nu} - \frac{1}{2} h_{\mu\nu} \bar R  &  =\kappa \bar T_{\mu\nu}-\frac12 h_{\mu\nu} \bar U(\Phi)
	\end{align}
	\begin{align}
	\label{EOM_scalar_field_P1}
	  \Phi\bar R &  -  (\Phi^2\,\bar U^(\Phi))^\prime =0,
	\end{align}
\end{subequations}
where the bar quantities are defined as follows: $\bar U(\Phi)=(\hat{R}\Phi-f(\hat{R}))/\Phi^2$, and the appropriate energy momentum tensor reads $\bar T_{\mu\nu}=\Phi^{-1}T_{\mu\nu}$.
One should also bear in mind that $\hat R_{\mu\nu}=\bar R_{\mu\nu}, \bar R= h^{\mu\nu}\bar R_{\mu\nu}=\Phi^{-1} \hat R$ and $h_{\mu\nu}\bar R=\ g_{\mu\nu}\hat R$. The above equations are field equations of the scalar-tensor action for the metric  $h_{\mu\nu }$ and (non-dynamical) scalar field $\Phi$.

Using these properties and the interpretation of the Palatini gravity it was shown \cite{aneta, anet} that the generalized Tolman-Oppenheimer-Volkoff (TOV) equations read\looseness=-1
\begin{align}
  \left(\frac{\Pi}{\Phi({r})^2}\right)'&=-\frac{G\tilde{A}\mathcal{M}}{r^2}\left(\frac{Q+\Pi}{\Phi({r})^2}\right)
  \left(1+\frac{4\pi r^3\frac{\Pi}{\Phi({r})^2}}{\mathcal{M}}\right)\label{tov_kon}\\ 
	\mathcal{M}(r)&= \int^r_0 4\pi \tilde{r}^2\frac{Q(\tilde{r})}{\Phi(\tilde{r})^2} d\tilde{r},\label{mass}
\end{align}
where the generalized energy density $Q$ and the pressure $\Pi$ are defined as
\begin{subequations}\label{defq}
 \begin{equation}
   \bar{Q}=\bar{\rho}+\frac{1}{2}\bar{U}=\frac{\rho}{\Phi^2}+\frac{U}{2\Phi^2}=\frac{Q}{\Phi^2},
 \end{equation}
\begin{equation}
  \bar{\Pi}=\bar{p}-\frac{1}{2}\bar{U}=\frac{p}{\Phi^2}-\frac{U}{2\Phi^2}=\frac{\Pi}{\Phi^2}.
\end{equation}
\end{subequations}
Recall that $\bar{U}$ and $\Phi$ depend on the choice of the model we are interested in. 

In what follows we will focus on the Starobinsky model $f(\hat R)=\hat R+\beta \hat R^2$ \cite{Starobinsky:1980te} considered in \cite{aneta2} in the context of the Lane-Emden equation. The generalized versions of Lane-Emden equation were already studied in \cite{aneta2, riazi, capH, saito, andre, sak, koy}, and some of them were obtained from the generalized TOV equations 
\cite{aneta, AltNS1, AltNS2, AltNS3, AltNS4, AltNS5, AltNS6, AltNS7, aw1}. 

We follow the Weinberg convention \cite{weinberg}, and hence the signature of the metric is $(-+++)$ while $\kappa=-8\pi G/c^4$.

\section{Exact solutions of the Lane-Emden equation}

The standard Lane-Emden equation coming from the Newtonian limit of the GR equilibrium equations (TOV equations) and polytropic equation of state $p=K\rho^\gamma$, where $K$ and $\gamma$ are polytropic parameters, has the following form:
\begin{equation}\label{le1}
    \frac{1}{\xi^2}\frac{d}{d\xi}\left(\xi^2\frac{d\theta}{d\xi}\right)+\theta^n=0;
\end{equation}
here $n=\frac{1}{\gamma-1}$. 

Recall the relation among dimensionless quantities appearing in (\ref{le1}) and the physical ones:
\begin{align}
 r&=r_c{\xi},\;\;\;\rho=\rho_c\theta^n,\;\;\;p=p_c\theta^{n+1},\label{def1}\\
 r^2_c&=\frac{(n+1)p_c}{4\pi G\rho^2_c},\label{def2}
\end{align}
where $p_c$ and $\rho_c$ are central pressure and energy density, respectively.

Equation (\ref{le1}) possesses three exact solutions for $n=\{0,1,5\}$ and, generally speaking, it proved pretty much  impossible to find exact solutions of the generalized Lane-Emden equations considered in \cite{riazi, capH, saito, andre, sak, koy}. Thus, for other values of $n$ or in the case of generalized Lane-Emden equations, the approximate and/or numerical methods were applied in order to examine the  properties of stars.\looseness=-1

As shown in \cite{aneta2}, the generalized Lane-Emden equation for the Starobinsky Lagrangian $f(\hat R)=\hat R+\beta \hat R^2$ in the Einstein frame is given by
\begin{equation}\label{LEo}
\frac{1}{\bar{\xi}}\frac{d^2}{d\bar{\xi}^2}\left(\left[1+\frac{2\alpha}{n+1}\theta^n\right]\bar{\xi}\theta\right)=-\theta^n,
\end{equation}
where $\alpha=\kappa c^2\beta\rho_c$, with $\rho_c$ being a central density. The extra term including $\alpha$ occurring
in (\ref{LEo}) comes from the potential $U$ appearing in (\ref{EOM_P1}) and definitions (\ref{defq}) whose form depends on the 
Lagrangian. The details of the derivation can be found in \cite{aneta2}.

Under the conformal transformation $\bar{\xi}^2=\Phi\xi^2$, where $\Phi=1+2\alpha\theta^n$, the above equation takes the  form\looseness=-1
\begin{equation}\label{LE}
\xi^2\theta^n\Phi+\displaystyle\frac{\Phi^{-1/2}}{1+\frac12 \xi\Phi_{\xi}/\Phi}\displaystyle\frac{d}{d\xi}\left(\displaystyle\frac{\xi^2\Phi^{3/2}}{1+\frac12 \xi\Phi_{\xi}/\Phi}\displaystyle\frac{d\theta}{d\xi} \right)=0.
\end{equation}
An approximation of \eqref{LE} was analyzed from the perspective of finding the numerical solutions in \cite{aneta2}. Notice that for $\alpha=0$ we have $\Phi=1$ and \eqref{LE} boils down to (\ref{le1}).

Now note that for $n=0$ equation (\ref{LEo}) becomes linear,
\begin{equation}\label{LE0}
\frac{1}{\bar{\xi}}\frac{d^2}{d\bar{\xi}^2}\left(\left[1+2\alpha\right]\bar{\xi}\theta\right)=-1,
\end{equation}
and its general solution is readily found to read
\begin{equation}\label{gso}
\theta=-\displaystyle\frac{\bar{\xi}^2}{6(1+2\alpha)}+C_0+\frac{C_1}{\bar{\xi}},
\end{equation}
where $C_i$ are arbitrary constants.

The transformation $\bar{\xi}^2=\Phi\xi^2$ in this case amounts to a rescaling $\bar{\xi}=(1+2\alpha)^{1/2}\xi$, so the general solution of (\ref{LE}) for $n=0$ reads
\begin{equation}\label{gs0o}
\theta=-\displaystyle\frac{\xi^2}{6}+C_0+\frac{C_1}{\xi},
\end{equation}
where we have rescaled $C_1$ for the sake of convenience.

If $C_1\neq 0$ then this solution has singularity at $\xi=0$, which is unphysical, so we should assume that $C_1=0$. Upon further imposing the boundary conditions in the Jordan frame, $\theta(0)=1$, $\theta'(0)=0$, we find that (\ref{gs0o}) becomes
\begin{equation}\label{gs0}
\theta=-\displaystyle\frac{\xi^2}{6}+1.
\end{equation}
Notice that it has the same form as the solution of the standard Lane-Emden equation (\ref{le1}) for $n=0$ describing 
incompressible stars, and for this reason it has been discussed in many textbooks (see e.g. \cite{weinberg}).

Our key observation now is that for $n=1$ the generalized Lane-Emden equation (\ref{LE}) has an exact solution
\begin{equation}\label{sol}
 \theta=\frac{15-\xi^2}{2\alpha(10+\xi^2)},
\end{equation}
which appears to be new and can be employed to describe a Newtonian neutron star. 

\section{Newtonian neutron star in Palatini gravity}

As we have already mentioned, the solutions of the Lane-Emden equation have to satisfy the boundary conditions, namely $\theta(0)=1$ and $\theta'(0)=0$. The solution (\ref{sol}) for $n=1$
turns out to satisfy them only if  $\alpha=\frac{3}{4}$. 

This value  lies in the allowed range $\alpha>-1/2$. Indeed, because of the boundary condition $\theta(0)=1$ the conformal factor $\Phi$ for $\xi=0$ takes the value $1+2\alpha$. Now, for $\alpha=-1/2$ this value becomes zero which is not allowed since the conformal factor may not change sign, cf.\ \cite{aneta2} for details. Moreover, the numerical analysis in \cite{aneta2} also indicates that the values of $\alpha$ smaller than $-1/2$ are unphysical.\looseness=-1 

This means that when we assume that the value of the central density of an average neutron star is $\rho_c\sim 8\cdot 10^{17}\frac{\text{kg}}{\text{m}^3}$, the Starobinsky parameter is found to be 
\begin{equation}
 \beta\sim-5.02978\cdot10^6\text{m}^2.
\end{equation}
Now we are able to find the star's mass, radius, and central density. Recall that by introducing
the dimensionless quantities \cite{edd, weinberg}:
\begin{align}
 \omega_n&:=-\xi^2\frac{d\theta}{d\xi}\mid_{\xi=\xi_R},\\
 \gamma_n&:=(4\pi)^\frac{1}{n-3}(n+1)^\frac{n}{3-n}\omega_n^\frac{n-1}{3-n}\xi_R,\\
 \delta_n&:=-\frac{\xi_R}{3\frac{d\theta}{d\xi}\mid_{\xi=\xi_R}}.
\end{align}
we can write down the star's mass, the mass-radius relation, central density, and temperature, in the form
\begin{align}
 M&=4\pi r_c^3\rho_c\omega_n,\label{mass2}\\
 R&=\gamma_n\left(\frac{K}{G}\right)^\frac{n}{3-n}M^\frac{n-1}{n-3}\xi_R,\\
 \rho_c&=\delta_n\left(\frac{3M}{4\pi R^3}\right),\\
 T&=\frac{K\mu}{k_B}\rho_c^\frac{1}{n}\theta,
\end{align}
where the last equation for the temperature profile is given by the assumption that the gas is ideal with the equation of state $T=\frac{k_B\rho T}{\mu}$, with $k_B$ and $\mu$
being the Boltzmann constant and mean molecular weight, respectively.\looseness=-1 

However, let us notice that the above 
quantities were defined for the General Relativity setting. Since we wish to apply them to the Palatini gravity, we must remember that the above expressions are the ones in Einstein frame. Thus, the quantities $\omega_n$
and $\delta_n$ should be rewritten; notice that the mass (\ref{mass}) is still written in the Einstein frame because the $r$-coordinate comes from the conformal metric $h$. This can be 
directly verified by applying the generalized Lane-Emden equation (\ref{LE}) to (\ref{mass}) while taking into account that  the conformal transformation in the case of Starobinsky model preserves the polytropic equation of state for small values of $p$ \cite{mana}. Thus, in the case of Palatini gravity we should have written
\begin{align}
 \omega_n&:=-\bar\xi^2\frac{d\theta}{d\bar\xi}\mid_{\xi=\xi_R},\label{J1}\\
 \delta_n&:=-\frac{\bar\xi_R}{3\frac{d\theta}{d\bar\xi}\mid_{\xi=\xi_R}}\label{J2}.
\end{align}
Then, applying the conformal transformation relation $\bar\xi^2=\Phi\xi^2$, we find that
\begin{align}
 \omega_n&=-\frac{\xi^2\Phi^\frac{3}{2}}{1+\frac{1}{2}\xi\frac{\Phi_\xi}{\Phi}}\frac{d\theta}{d\xi}\mid_{\xi=\xi_R},\label{omega}\\
 \delta_n&=-\frac{\xi_R}{3\frac{\Phi^{-\frac{1}{2}}}{1+\frac{1}{2}\xi\frac{\Phi_\xi}{\Phi}}\frac{d\theta}{d\xi}\mid_{\xi=\xi_R}}.
\end{align}
Now we are able to calculate the physical values of the star's mass.
Moreover, since the solution for $n=1$ in the case of the standard Lane-Emden
equation is known, $\theta_{N}=\frac{\sin{\xi}}{\xi}$, we can compare Newtonian neutron stars in the two models. Here and below we will denote the solution and values obtained from the standard Lane-Emden equation by the subscript $N$. 

\begin{figure}[h!t]
\centering
\includegraphics[scale=.6]{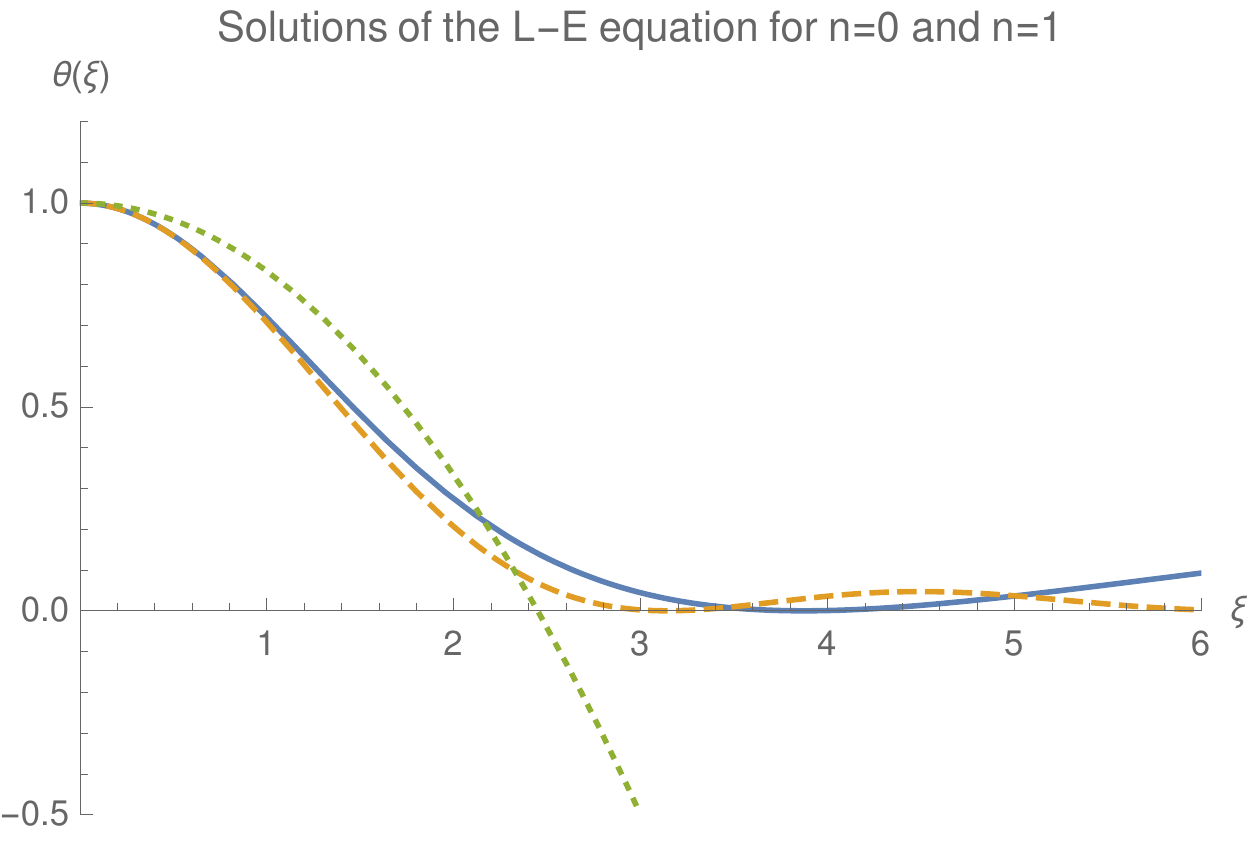}
\caption{(color online) Plots of the exact solutions for the case $n=1$. The dashed curve represents the solution $\theta=\frac{\sin{\xi}}{\xi}$
of the standard equation while the continuous curve is the solution of the Palatini one, $\theta=\frac{15-\xi^2}{2\alpha(10+\xi^2)}$, with
$\alpha=3/4$. The solution for $n=0$, which is the ame for both cases, is given by the dotted curve.}
\label{fig1}
\end{figure}

We plot the solutions in Figure \ref{fig1}. Notice that Figure \ref{fig1} also represents the dimensionless temperature profile as $T\sim\theta$. 

The neutron star's radius $\xi_R$ is defined by the first zero of the solution, $\theta(\xi_R)=0$, and hence we find 
that $\xi_R=\sqrt{15}$ while
 $\omega_1\sim 3.87$, and $\delta_1=5$. Therefore, we immediately can compare the masses and radii of the stars 
\begin{align}
 \frac{M}{M_N}=\frac{\omega_1}{\omega_N}\sim 1.23,\;\;\frac{R}{R_N}\sim 1.23,\;\;\frac{\rho}{\rho_N}\sim 0.99
\end{align}
where we have used $(\xi_R)_N=\pi$ and $\omega_N=\pi$ \cite{weinberg}.

\section{Conclusions}
We have found two exact solutions for $n=0$ and $n=1$ of the modified Line-Emden equation (\ref{LE}) using the Einstein-Jordan frame correspondence
\cite{Afonso:2018bpv,Afonso:2018mxn,beltran,olek}, keeping in 
mind that physical properties of the star such as mass and radius are the ones appearing in Jordan frame, which we demonstrated clearly in the discussion 
preceding the formulas (\ref{J1}) and (\ref{J2}). We have not examined the result for $n=0$ since it would describe an incompressible star. We should notice that 
the latter solution has exactly the same form as the solution of the standard Lane-Emden equation. However, the case of the Newtonian neutron star (the solution for $n=1$), although being just a toy model, 
show that one deals with the stellar objects bigger than the stars of the pure GR models. Let us also stress that in this work we have used the exact formula for obtaining the star's mass, that is, the equations (\ref{mass2}) and (\ref{omega}) in contrast to the previous work \cite{aneta2} where the approximated formula for mass was used.

We expect that the examination of the full relativistic theory,
described by the modified TOV equations (\ref{tov_kon}) and (\ref{mass}), will also provide a similar conclusion, that is, that the Palatini star is larger and heavier for the negative Starobinsky parameter (positive $\alpha$ parameter), without 
a need of introducing exotic matter in their internal structure. It should be clarified here that the result was obtained for the negative Starobinsky parameter 
$\beta$: the parameter $\alpha$ appearing in the modified Lane-Emden equation consists of the constant $\kappa$, which had been defined to be negative.

Moreover, requiring that the solution (\ref{sol}) has to satisfy the boundary conditions accompanying the Lane-Emden equation, we obtained the exact value of the parameter 
$\alpha=\frac{3}{4}$ which led to the fact that the Starobinsky parameter is negative and of the order $10^6  \text{m}^2$ under the assumption of the central energy density $\rho_c\sim 8\cdot 10^{17}\frac{\text{kg}}{\text{m}^3}$.

Let us also mention a discussion in \cite{beltran} that in the Palatini theories (for example, $f(\hat R)$ gravity and Born-Infeld 
inspired gravity,  cf.\ e.g.\ \cite{b1,b2,b3,b4}) 
the energy density must be continuous and differentiable function in order to avoid  divergences in the field equations. This 
comes from the fact that one obtains $\hat R=\hat R(\rho)$ which appears in the conformal function $\Phi$ taking part in the  conformal transformation which includes the derivatives of $\rho$. 

Therefore, we would like to conclude that the Palatini gravity is a viable alternative to General Relativity since it passes Solar
System tests (vacuum solutions are equivalent to General Relativity with cosmological constant) \cite{barraco, al1, al2}, 
introduces inflation preserving late time accelerated expansion \cite{BSSW1,BSSW2,BSS,SSB}, gives clues on the 
Dark Matter problem \cite{wojn_galax}, and provides conditions for existence of stable relativistic stars \cite{anet} that are similar to General
Relativity.

\acknowledgments{AW acknowledges financial support from FAPES (Brazil). The research of AS was supported in part by the Grant Agency of the Czech
Republic (GA \v{C}R) under grant P201/12/G028 and under RVO funding for I\v{C}47813059.

AS is pleased to thank the Institute for Theoretical Physics of the Wroc\l aw University and, in particular, Andrzej Borowiec, for the warm hospitality extended to him in the course of his visits to Wroc\l aw where the present research project was initiated.

Furthermore, the authors warmly thank Andrzej Borowiec, Gonzalo Olmo, Diego Rubiera-Garcia, and Hermano Velten for stimulating discussions and comments.}

\end{document}